\documentclass[aps, twocolumn, pra, showpacs]{revtex4}
\usepackage{graphicx}
\usepackage{amsmath}
\usepackage{amsfonts}
\usepackage{amssymb}
\usepackage{epsfig}

\begin{document}

\title{Entropy majorization, thermal adiabatic theorem, and quantum phase
transitions}
\author{Shi-Jian Gu}
\email{sjgu@phy.cuhk.edu.hk}
\affiliation{Department of Physics and ITP, The Chinese University of Hong Kong, Hong
Kong, China}

\begin{abstract}
Let a general quantum many-body system at a low temperature adiabatically cross
through the vicinity of the system's quantum critical point. We show that the
system's temperature is significantly suppressed due to both the entropy
majorization theorem in quantum information science and the entropy
conservation law in adiabatic processes. We take the one-dimensional
transverse-field Ising model and spinless fermion system as concrete examples
to show that the inverse temperature might become divergent around their
critical points. Since the temperature is a measurable quantity in experiments,
our work, therefore, provides a practicable proposal to detect quantum phase
transitions in view of both quantum information science and statistical
physics.
\end{abstract}

\date{\today }
\maketitle




\section{Introduction}

In the recent years, quantum phase transitions \cite{Sachdev} have been paid
considerable attentions within various fields of physics because they can
provide valuable information about novel types of matter that emerge in the
vicinity of the absolute zero temperature. Examples that have been studied
both theoretically and experimentally include high-$T_c$ superconductors,
fractional quantum Hall liquids, quantum magnets, Mott-insulators etc.
Unlike thermal phase transitions that occur at finite temperatures and are
driven by thermal fluctuations, quantum phase transitions occur at the
absolute zero temperature and are driven solely by quantum fluctuations.
Conventionally, quantum phase transitions are characterized by singularities
of the ground-state energy. The order of transitions is defined by
discontinuities (or singularities) occurring in the $n$th-order derivative
of the energy. For instance, the second-order quantum phase transition
corresponds to the discontinuity (or singularities) in the second derivative
of the energy. Such a characterization is intrinsically inherited from
traditional studies on thermal phase transitions since the
finite-temperature free energy becomes the ground-state energy in the limit
of zero temperature. Theoretical frameworks, such as Landau-Ginzburg-Wilson
spontaneous symmetry-breaking theory, have been used to understand
continuous quantum phase transitions too. In Landau's paradigm, local order
operators and their correlation functions play a central role in studies on
quantum phase transitions. The nonvanishing value of the order parameter at
long distance characterizes a symmetry-breaking phase, a unique feature
which exists only for a system with infinite degrees of freedom.

Recently, a huge interest was raised in the attempt of characterizing
quantum phase transitions in terms of the concepts borrowed from quantum
information science \cite{Nielsen1}. Due to the mutual relation between
fluctuations and correlations, people are curious about the role of
entanglement, a pure quantum correlation existing uniquely in quantum
systems, in those transitions driven solely by quantum fluctuations.
Hundreds of work (For a review, see Ref. \cite{LAmico08}) have been
published since the two original works finished by Osterloh \emph{et al}
\cite{AOsterloh2002}, and Osborne and Nielsen \cite{TJOsbornee},
respectively. Though a unified theory on the role of entanglement in quantum
phase transitions is still unavailable, some definitive conclusions have
been commonly accepted \cite{LAmico08}. Another promising approach to
quantum phase transitions is based on quantum fidelity, a concept emerging
also in quantum information science. The fidelity measures the similarity
between two ground states, hence is excepted to show a dramatic change
across the transition points \cite{Zanardi06,HQZhou0701,WLYou07}. This
motivated people to start exploring its role played in quantum phase
transitions (For a review, see Ref. \cite{SJGUreview}). Moreover, as the
fidelity is a space geometrical quantity, no a priori knowledge of the order
parameter and symmetry breaking of the system is assumed, it is a great
advantage to study quantum phase transitions using the fidelity approach.

Experimentally, however, to study the ground-state entanglement and fidelity
in quantum many-body systems is still a challenging problem. The
nuclear-magnetic-resonance quantum simulator, though is promising, can only
measure entanglement \cite{XPengEnt} and the fidelity \cite%
{JZhangPRL100501,JZhang08081536} in few-body systems (for instance spin
dimer). The critical phenomena, such as scaling properties and various
critical exponents, are not able to be studied in experiments yet.

In this paper, we propose a scheme to study quantum criticality based on the
entropy majorization theorem \cite{MANielsen99}, an important theorem in
quantum information science. Our motivation originates from three facts:

\begin{enumerate}
\item A true quantum phase transition occurs at the absolute zero
temperature, which, however, cannot be reached rigorously in experiments. To
understand the ground-state properties, experiments are usually done at low
or untra-low temperatures. Therefore, it is useful to develop a proposal to
study quantum phase transitions by considering the effects of temperature.

\item The ground state cannot be isolated from the corresponding low-lying
excitations also. From this point of view, the change in the low-energy
spectra \cite{GSTian2003} plays an important role for understanding quantum
critical phenomena.

\item At low temperatures, the entropy majorization theorem manifests that
the entropy is significantly enhanced around the critical point due to a
higher density of state nearby. While the entropy is also required to be
conserved in adiabatic processes because there is not heat transfer during
the process. In order to satisfy the entropy conservation law in the
adiabatic process, the temperature has to be suppressed and the \emph{%
adiabatic inverse temperature} (denoting the inverse temperature in
adiabatic processes) might even become singular at the critical point.
\end{enumerate}

Our general motivation will be proved via the entropy majorization theorem,
and then confirmed by concrete examples, including a simple two-level
system, the one-dimensional transverse-field Ising model, and the
one-dimensional spinless fermion system.

This work is organized as follows. In section \ref{sec:warm-up}, we give a
warm-up example to show explicitly that the adiabatic inverse temperature
will become singular when a ground-state level-crossing occurs. In section %
\ref{sec:major}, we show that, based on the entropy majorization theorem,
the significant change in the low-energy spectra might lead to a singular
adiabatic inverse temperature around the quantum critical point. In section %
\ref{sec:ising}, we take the one-dimensional transverse-field Ising model as
an example to show that the adiabatic inverse temperature becomes singular
around the critical point. In section \ref{sec:fermions}, we study the
thermodynamics of the one-dimensional spinless fermion system at low
temperatures, and show explicitly that the adiabatic inverse temperature is
divergent at the quantum phase transition point. In section \ref{sec:diss},
we discuss potential application of our motivation to experiments. Finally,
we give a summary in section \ref{sec:sum}.

\section{A Warm-up two-level model}

\label{sec:warm-up}

\begin{figure}[tbp]
\includegraphics[width=8cm]{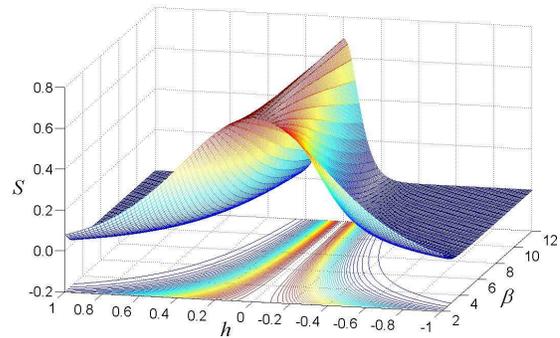}
\caption{(Color online) The thermal entropy of the two-level system as a
function of $h$ and $\protect\beta$. Contour lines in the $h-\protect\beta$
plane denote the adiabatic evolution paths.}
\label{fig:2levelentropy.eps}
\end{figure}

To begin with, we first have a look at a two-level toy model describing a
free spin in an external field as a warm-up example. The model's Hamiltonian
reads as
\begin{equation}
H=-h\sigma ^{z},
\end{equation}%
where $\sigma ^{z}$ is the $z$-component of Pauli matrices $\{\sigma
^{x},\sigma ^{y},\sigma ^{z}\}$. In the 1/2 spin basis $\left\{ \left\vert
\uparrow \right\rangle ,\left\vert \downarrow \right\rangle \right\} $, the
Hamiltonian is already diagonal and the corresponding eigenenergies are
\begin{equation}
E_{1}=-h,E_{2}=h.
\end{equation}%
Therefore, if $h>0$, the ground state is $\left\vert \uparrow \right\rangle $%
; while if $h<0$, the ground state becomes $\left\vert \downarrow
\right\rangle $. A simple first-order quantum phase transition induced by
the ground-state level-crossing occurs at the transition point $h_{c}=0$.

The quantum phase transition occurring in such a two-level system looks
rather trivial. However, it captures one of the most important properties in
quantum critical phenomena, a vanishing energy gap (or more generally a
significant change in the density of state). Suppose the absolute zero
temperature is unavailable, the thermal equilibrium state, according to
quantum statistical mechanics can only be described by a thermal density
matrix. For the present system, the thermal density matrix at a temperature $%
\beta =1/T$, with the Boltzmann constant $k_{B}=1$, is
\begin{equation}
\rho =\frac{1}{Z}\left(
\begin{array}{cc}
e^{-\beta h} &  \\
& e^{\beta h}%
\end{array}%
\right)
\end{equation}%
where%
\begin{equation}
Z=e^{\beta h}+e^{-\beta h}
\end{equation}%
is the partition function. From the thermal density matrix, we can see that
the probability of two states are%
\begin{equation}
p_{1}=\frac{e^{\beta h}}{Z},\text{ \ }p_{2}=\frac{e^{-\beta h}}{Z},
\end{equation}%
respectively. Therefore, the entropy of the system reads
\begin{eqnarray}
S &=&-p_{1}\ln p_{1}-p_{2}\ln p_{2} \\
&=&\ln (1+e^{-2\beta h})+\frac{2\beta he^{-2\beta h}}{1+e^{-2\beta h}}.
\notag
\end{eqnarray}%
Now we suppose the system adiabatically envolves from $h=h_{0}$ and $\beta
=\beta _{0}$ to $h=-h_{0}$ with $h_{0}>0$. During the adiabatic process, the
system's entropy is required to be conserved, i.e. $S=\mathrm{const}$, hence%
\begin{equation}
\ln (1+e^{-2\beta h})+\frac{2\beta he^{-2\beta h}}{1+e^{-2\beta h}}=\mathrm{%
const.}
\end{equation}%
In this expression, the only relevant parameter that changes the entropy is $%
\beta h$. Therefore, to fix the entropy, we need only to fix $\beta h$, i.e.
\begin{equation}
h\beta =h_{0}\beta _{0},
\end{equation}%
such that
\begin{equation}
\beta =\frac{h_{0}\beta _{0}}{h}.
\end{equation}%
We can see clearly that the adiabatic inverse temperature $\beta $ becomes
divergent as $h$ tends to zero. That is, if we change the external field
slowly enough, the temperature will become zero at the transition point.

As a numerical demonstration, we show, in Fig. \ref{fig:2levelentropy.eps},
the thermal entropy of the two-level system as a function of $h$ and $\beta $%
, and its contour map on the $h-\beta $ plane. The lines on the contour map
denote adiabatic paths along which the system evolves, from which we can
judge that the adiabatic inverse temperature is divergent around $h=0$.

\section{Entropy majorization theorem and quantum criticality}

\label{sec:major}

The discussion in the above two-level system shows a clear picture that the
change in energy spectra around the critical point might lead to a divergent
adiabatic inverse temperature. In this section, we want to generalize this
picture to an arbitrary quantum many-body system.

We consider a general quantum many-body system described by the Hamiltonian
\begin{equation}
H=H_{0}+\lambda H_{I},  \label{eq:Hamiltonian}
\end{equation}%
where $H_{I}$ is the driving Hamiltonian and $\lambda $ denotes its
strength. Without loss of generality, a quantum phase transition is supposed
to occur in the system's ground state $|\phi _{0}(\lambda )\rangle $ at the
critical $\lambda _{c}$. Since there is no rigorous ground state in
experiments, the thermal equilibrium state of the system at a high $\beta$
(hence low $T$), according quantum statistical physics, can be expressed as
a density matrix
\begin{equation}
\rho (\beta ,\lambda )=\frac{1}{Z}\sum\limits_{n}{e^{-\beta E_{n}}\left\vert
\phi _{n}\right\rangle \left\langle \phi _{n}\right\vert .}
\end{equation}%
Here $|\phi _{n}(\lambda )\rangle $ which satisfy%
\begin{equation}
H(\lambda )|\phi _{n}(\lambda )\rangle =E_{n}(\lambda )|\phi _{n}(\lambda
)\rangle ,
\end{equation}%
define a set of orthogonal complete basis, and
\begin{equation}
Z=\sum_{n}{e^{-\beta E_{n}}}
\end{equation}%
is the partition function. The entropy of the system then can be expressed as%
\begin{equation}
S=\beta \left( \left\langle E\right\rangle -F\right) =-\sum_{n}p_{n}\ln
p_{n}.
\end{equation}%
Here $\left\langle E\right\rangle $ and $F$ are the internal energy and the
free energy respecticaly, and
\begin{equation}
p_{n}=\frac{1}{\bar{Z}}e^{-\beta \left( E_{n}-E_{0}\right) },
\end{equation}%
where $\bar{Z}=Ze^{\beta E_{0}}$ and with%
\begin{eqnarray}
\sum p_{n} &=&1, \\
p_{0} &\geqslant &p_{1}\geqslant p_{2}\cdots .
\end{eqnarray}%
$p_{i}=p_{j}$ occurs iff $E_{i}=E_{j}$.

In the thermodynamic limit, the entropy of the system can be written as%
\begin{equation}
S=-N\int p(\epsilon )\log \left[ p(\epsilon )\right] \rho (\epsilon
)d\epsilon ,
\end{equation}%
where $\epsilon =E/N$ with $N$ being the system size, and $\rho (\epsilon )$
denotes the density of state at $\epsilon $. Now we are going to show that a
significant change in the density of state around the ground state will
change the temperature or the entropy, depending on which one keeps
constant. The change of density of state includes but not limit to a
vanishing gap around the critical point \cite{Sachdev} or emergence of a Van
Hove singularity at the ground state \cite{VanHove}.

\subsection{Quantum phase transitions induced by a vanishing gap}

We consider the system at two points $\lambda $ and $\lambda ^{\prime }$ in
the parameter space. Without loss of generality, we assume that $\lambda
^{\prime }$ is closer to the critical point $\lambda _{c}$, i.e. $|\lambda
^{\prime }-\lambda _{c}|<|\lambda -\lambda _{c}|$, and the temperature is
the same. So there are two thermal probability distributions $\left\{
p_{n}\right\} $ and $\left\{ p_{n}^{\prime }\right\} $ at $\lambda $ and $%
\lambda ^{\prime }$, respectively. The low-temperature properties of the
system are determined by both the ground state and the low-lying
excitations, we need to consider only their contributions to the entropy of
the system. Since the system is gapped and the gap is assumed to vanish at
the critical point, we have $E_{1}^{\prime }-E_{0}^{\prime }<E_{1}-E_{0}$.
Defining
\begin{equation}
\bar{Z}_{k}=\sum_{n=0}^{k}e^{-\beta \left( E_{n}-E_{0}\right) },\text{ \ }%
\bar{Z}_{k}^{\prime }=\sum_{n=0}^{k}e^{-\beta \left( E_{n}^{\prime
}-E_{0}^{\prime }\right) },
\end{equation}%
we have%
\begin{eqnarray}
\bar{Z}_{0} &=&\bar{Z}_{0}^{\prime }, \\
\bar{Z}_{1} &<&\bar{Z}_{1}^{\prime },  \notag \\
&&\vdots  \notag \\
\bar{Z}_{\Lambda } &<&\bar{Z}_{\Lambda }^{\prime }.  \notag
\end{eqnarray}
where $\Lambda $ is a cutoff. Therefore, there exists a range in the energy
space, in which that%
\begin{eqnarray}
p_{0} &>&p_{0}^{\prime }, \\
p_{1} &<&p_{1}^{\prime },  \notag \\
&&\vdots  \notag \\
p_{\Lambda } &<&p_{\Lambda }^{\prime }  \notag
\end{eqnarray}%
such that
\begin{equation}
\sum_{j}^{k}p_{j}>\sum_{j}^{k}p_{j}^{\prime },
\end{equation}%
for $0<k\leq \Lambda $. Here $\Lambda $ can be finite or even as large as
the dimension of the whole Hilbert space depending on both the temperature
and the normalization condition of the Boltzmann distribution. If the
temperature is low enough, contributions from those eigenstates above $%
E_{\Lambda }$ are neglectable. Therefore, in such a region, we can say the
distribution $\left\{ p_{n}\right\} $ majorizes $\left\{ p_{n}^{\prime
}\right\} $ \cite{BCArnoldbook}. Since $\left\{ p_{n}\right\} $ and $\left\{
p_{n}^{\prime }\right\} $ denote the probability distribution, we have,
according to the entropy majorization theorem \cite{MANielsen99}
\begin{equation}
S(p)<S(p^{\prime }).
\end{equation}%
The inequality manifests that the entropy increases as the energy gap
vanishes.

\begin{figure}[tbp]
\includegraphics[width=8cm]{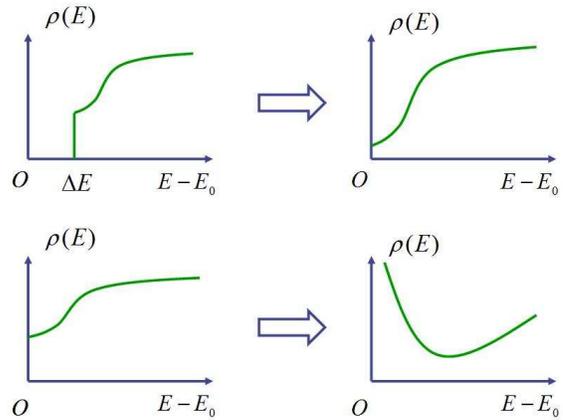}
\caption{(Color online) Possible changes of density of state when quantum
phase transition occur. Upper: A gap vanishes the system tends a critical
point. A typical case is the one-dimensional transverse-field Ising model.
Lower: The density of state in low-energy region becomes singular around the
critical point. A typical case is the the one-dimensional spinless fermion
system. }
\label{fig:dos.eps}
\end{figure}

On the other hand, the thermal entropy is a monotonically increasing
function of the temperature, to ensure the entropy to be conserved during
adiabatic processes, the temperature $T$ at $\lambda ^{\prime }$ must be
lower than that at $\lambda $, hence the adiabatic inverse temperature $%
\beta $ increases. To see the singularity of the adiabatic inverse
temperature, the starting temperature before the adiabatic evolution must be
low enough such that the Boltzmann probability of the ground state $1/\bar{Z}
$ is finite, say 1/10 which does not vanish even in the thermodynamic limit.
In this case, the system's entropy usually is intensive given there is no
singularity in the density of state below $E_\Lambda$. Then if the energy
gap is closed, the density of state becomes continuous around the ground
state (see top right of Fig. \ref{fig:dos.eps}), the adiabatic inverse
temperature becomes divergent. Otherwise the entropy would be extensive
(divergent as $\sim N$) if $\beta $ is still finite.

\subsection{Quantum phase transitions induced by a Van Hove singularity}

Generally, the significant change in the density of state is not limited to
the case of a vanishing gap. A straightforward example is the emergence of a
Van Hove singularity at the ground state (see bottom right of Fig. \ref%
{fig:dos.eps}). This case usually occurs as the low-lying excitations'
dispersion is, though gapless, intrinsically changed around the critical
point, for instance from linear to quadratic.

Assuming at $\lambda $, the density of state is a continuous function in the
vicinity of the ground state, then$,$%
\begin{equation}
S=-N\int_{0}^{\Lambda }p(\epsilon )\ln \left[ p(\epsilon )\right] \rho
(\epsilon )d\epsilon  \label{eq:vonentropy1}
\end{equation}%
where $\Lambda $ is a cut-off due to $p(\epsilon )\sim e^{-\beta N\epsilon
}/Z$; and at another point $\lambda ^{\prime }$ closer to the critical
point, the entropy becomes%
\begin{equation}
S^{\prime }=-N\int_{0}^{\Lambda }p(\epsilon )\ln \left[ p(\epsilon )\right]
\rho ^{\prime }(\epsilon )d\epsilon .  \label{eq:vonentropy2}
\end{equation}%
Since $\rho (\epsilon =0)$ becomes more and more singular as $\lambda $
tends to $\lambda _{c}$, the integrand in Eq. \ref{eq:vonentropy1} majorizes
that of Eq. (\ref{eq:vonentropy2}) due to that
\begin{equation}
Z=N\int_{0}^{\Lambda }e^{-\beta N\epsilon }\rho (\epsilon )d\epsilon .
\end{equation}%
Therefore, according to the entropy majorization theorem, we again have $%
S<S^{\prime }$. To ensure the entropy conservation during adiabatic
processes, the adiabatic inverse temperature in Eq. (\ref{eq:vonentropy2})
must be enhanced. Moreover, if $\rho (\epsilon =0)$ is divergent at the
critical point, $\beta$ is expected to be divergent too.

\subsection{Potential scaling issues based on the Landau-Zener model}

Quantum phase transitions occurring in few-body systems are usually induced
by the ground-state level-crossing. Nevertheless, most interesting critical
phenomena occur in thermodynamic systems where the singularity is due to the
infinite degree of freedom in the thermodynamic limit. That is, in these
cases, the ground-state properties do not show any singular behavior for a
finite sample. Then the scaling analysis becomes very important. Here we now
take the Landau-Zener model \cite{LDLandaubook,CZener} to illustrate the
such a picture. The model Hamiltonian reads as%
\begin{equation}
H=\left(
\begin{array}{cc}
\lambda & \omega (N) \\
\omega (N) & -\lambda%
\end{array}%
\right) ,
\end{equation}%
in the basis $\{\left\vert \uparrow \right\rangle ,\left\vert \downarrow
\right\rangle \}$, where $\omega (N)$ denotes $N$-dependent hoping amplitute
between two levels. The eigenstates of the Hamiltonian are%
\begin{eqnarray}
\psi _{1} &=&\cos \frac{\theta }{2}\left\vert \uparrow \right\rangle +\sin
\frac{\theta }{2}\left\vert \downarrow \right\rangle \\
\psi _{2} &=&-\sin \frac{\theta }{2}\left\vert \uparrow \right\rangle +\cos
\frac{\theta }{2}\left\vert \downarrow \right\rangle  \notag
\end{eqnarray}%
with $\cos \theta =\varepsilon /\sqrt{1+\varepsilon ^{2}},\sin \theta =1/%
\sqrt{1+\varepsilon ^{2}}$, and $\varepsilon =\lambda /\omega $. The energy
gap between two states are $\Delta E=2\sqrt{\omega ^{2}+\lambda ^{2}}$. If $%
\omega (N)$ is finite, the energy gap is always finite. This case
corresponds to a finite system. However, if $\omega (N)$ vanishes in the
infinite $N$ limit, for instance $\omega (N)\sim 1/N^{\mu }$ with $\mu >0$,
a quantum phase transition then occurs at $\lambda =0$.

We let the system adiabatically envolve from $\lambda _{0}$ and at
temperature $\beta _{0}$. The adiabatic process requires that
\begin{equation}
\beta \sqrt{\omega ^{2}+\lambda ^{2}}=\beta _{0}\sqrt{\omega ^{2}+\lambda
_{0}^{2}},
\end{equation}%
hence%
\begin{equation}
\beta =\frac{\beta _{0}\sqrt{\omega ^{2}+\lambda _{0}^{2}}}{\sqrt{\omega
^{2}+\lambda ^{2}}}
\end{equation}%
which reaches a maximum (for a finite $N$) at $\lambda =0$. In the large $N$
limit, we have
\begin{equation}
\beta \sim N^{\mu }
\end{equation}%
at $\lambda =0$. From the above relation, we can see that the adiabatic
inverse temperature diverges as the inverse of energy gap according to the
Landau-Zener model.

\section{The one-dimensional transverse-field Ising model}

\label{sec:ising}

\begin{figure}[tbp]
\includegraphics[width=8cm]{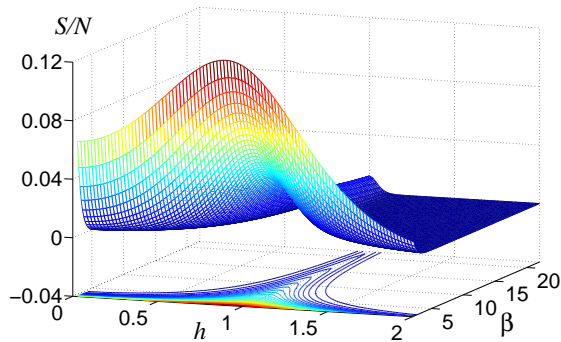}
\caption{(Color online) The thermal entropy per site of the one-dimensional
transverse-field Ising model as a function of $h$ and $\protect\beta$.
Contour lines in the $h-\protect\beta$ plane denote the adiabatic evolution
paths.}
\label{fig:ising.eps}
\end{figure}

In this section, we consider a continuous quantum phase transition occurring
in the ground state of the one-dimensional transverse-field Ising model.
Such a phase transition is believed to be well characterized by the
Landau-Ginzburg-Wilson spontaneous symmetry-breaking paradigm. Moreover, due
to its simplicity and exact solvability, the one-dimensional
transverse-field Ising model becomes one of the most popular models for
studying quantum phase transitions. The model Hamiltonian reads as%
\begin{eqnarray}
H &=&-\sum_{j=1}^{N}\left( \sigma _{j}^{x}\sigma _{j+1}^{x}+h\sigma
_{j}^{z}\right) ,  \label{eq:Hamiltonian_Ising0} \\
\sigma _{1}^{x} &=&\sigma _{N+1}^{x},
\end{eqnarray}%
where $h$ is the transverse field and $N$ is the number of spins. As implied
from the model's name, the Hamiltonian describes a chain of spins with the
nearest-neighboring Ising interaction along $x$-direction, and all spins are
subject to a transverse magnetic field $h$ along $z$-direction. If $h$ is
infinite, all spins are polarized along $z$-direction, the ground state is
paramagnetic in $x$-direction. While if $h=0$, the system becomes a
classical Ising chain whose ground state is ferromagnetic and presented a
true long-range order in $x$-direction. A quantum phase transition from a
ferromagnetic phase to a paramagnetic phase occurs at the critical point $%
h_{c}=1$.

The Ising model can be diagonalized exactly through three transformations
together, i.e. the Jordan-Wigner transformation, Fourier transformation, and
Bogoliubov transformation. The Hamiltonian finally becomes a quasi-free
fermion system in momentum space,
\begin{equation}
H=\sum_{k}\epsilon (k)\left( 2b_{k}^{\dagger }b_{k}-1\right) ,
\end{equation}%
where%
\begin{equation}
\epsilon (k)=\sqrt{1-2h\cos (k)+h^{2}}
\end{equation}%
is the dispersion relation of the quasi particles defined by $b_{k}^{\dagger
}$. The dispersion relation shows that, in the thermodynamic limit, the
system is gapless only at $h=1$, and gapped in both phases of $0<h<1$ and $%
h>1$.

According to quantum statistical mechanics, the partition function of the
system at $\beta $ can be calculated as
\begin{equation}
Z=\prod_{k}e^{\beta \epsilon (k)}\left[ 1+e^{-2\beta \epsilon (k)}\right] .
\end{equation}%
Then according to thermodynamics, $F=-T\ln Z$ and $S=-\partial F/\partial T$%
, the free energy and the thermal entropy are
\begin{eqnarray}
F &=&E_{0}-\frac{NT}{2\pi }\int_{-\pi }^{\pi }\ln \left[ 1+e^{-2\beta
\epsilon (k)}\right] dk \\
S &=&\frac{N}{2\pi }\int_{-\pi }^{\pi }\left( \ln \left[ 1+e^{-2\beta
\epsilon (k)}\right] +\frac{2\beta \epsilon (k)}{1+e^{2\beta \epsilon (k)}}%
\right) dk,  \notag \\
&&
\end{eqnarray}%
where $E_{0}$ is the ground-state energy.

It is not easy to get the final expression of the entropy explicitly due to
the presence of the energy gap. Nevertheless, we can calculate the system's
entropy as a function of $h$ and $\beta $ numerically. The results are shown
in Fig. \ref{fig:ising.eps}. From the figure, we first can see that the
thermal entropy is a monotonically decreasing function of $\beta$, this
observation is consistent with thermodynamics. Secondly, the thermal entropy
decreases more quickly in the both gapped phase, while rather slowly around
the critical point. Such a difference is due to the gap protection in the
non-critical region. Thirdly, if we fix $\beta$, we can see that the entropy
increases significantly at the transition point. The physics behind the
phenomenon is the entropy majorization theorem as we discussed in section %
\ref{sec:major}. Finally, that of significant importance is the contour map
of the thermal entropy on the $h-\beta$ plane. Clearly, a single line on the
contour map denotes an isoline of the thermal entropy. If the transverse
field is changed slowly enough, these lines are just paths along which that
the system evolves. Then if the initial inverse temperature is comparable
with the inverse energy gap, the inverse temperature becomes divergent at
the critical point.

\section{The one-dimensional spinless fermion system}

\label{sec:fermions}

In this section, we consider a quantum phase transition induced by
continuous level-crossing, as illustrated by the one-dimensional spinless
fermion system subject to a chemical potential. Such a phase transition is
basically different from that occurring in the one-dimensional
transverse-field Ising model. The later is due to the avoided
level-crossing, while the former is caused by the continuous ground-state
level-crossing. The model Hamiltonian can be written as
\begin{equation}
H=-t\sum_{j}\left( c_{j+1}^{\dag }c_{j}+c_{j}^{\dag }c_{j+1}\right) +\mu
\sum_{j}c_{j}^{\dag }c_{j},
\end{equation}%
where $c_{j}^{\dag }$ and $c_{j}$ are creation and annihilation operators
for fermions at site $j$, $t$ is the hoping integral, and $\mu $ is the
chemical potential. The Hamiltonian can be diagonalized via a simple Fourier
transformation. In the momentum space, its reads as%
\begin{equation}
H=\sum_{k}\left( \mu -2t\cos k\right) c_{k}^{\dag }c_{k}-\frac{N\mu }{2},
\end{equation}%
where $N=\sum_{j}c_{j}^{\dag }c_{j}$. Besides the global energy shift $N\mu
/2$, the ground state of the system is determined by the chemical potential $%
\mu $. If $\mu <\mu _{c}(=2t)$, the ground state of the system is partially
filled with fermions, the Fermi point is nonzero and determined by $\cos
k_{F}=\mu /2t$. In this case, the dispersion of the low-lying excitations is
linear without any gap. If $\mu >\mu _{c}$, the ground state becomes a
vacuum state with no fermions. Moreover, a finite energy will be introduced
if one add a fermion to the ground state. A quantum phase transition occurs
at the $\mu =\mu _{c}$ at which the ground state is still gapless, while the
dispersion of the low-lying excitations is quadratic. The transition of the
second order.

The partition function of the system can be calculated as%
\begin{equation}
Z=z^{-N/2}\prod_{k}\left[ 1+z\exp \left( 2t\beta \cos k\right) \right] ,
\end{equation}%
where $z=\exp (\mu \beta )$. According to thermodynamics, the free energy
can be calculated as
\begin{equation}
F=-N\mu -\frac{N}{2\pi \beta }\int_{-\pi }^{\pi }\ln \left[ 1+\exp \left(
-\beta (\mu -2t\cos k)\right) \right] dk
\end{equation}%
and the entropy%
\begin{eqnarray}
S &=&\frac{N}{2\pi }\int_{-\pi }^{\pi }u\left[ \beta \left( h-2t\cos
k\right) \right] dk, \\
u(x) &=&\ln \left( 2\cosh \frac{x}{2}\right) -\frac{x}{2}\tanh \frac{x}{2}.
\end{eqnarray}%
At low temperatures, the integration range in $x$ space become infinite, so
the entroy can be evaluate as%
\begin{equation}
S\simeq \frac{\pi N}{6t\beta \cos ^{-1}(\mu /2t)}.
\end{equation}%
Therefore, in the adiabatic process, if the entropy is conserved, then%
\begin{equation}
\beta \simeq \frac{\pi N}{6tS\cos ^{-1}(\mu /2t)}
\end{equation}%
which diverges at the critical point $\mu _{c}=2t$. When $\mu \rightarrow
\mu _{c}^{-}$, the ground state keeps gapless so the transition has nothing
to do with a vanishing gap. A careful scrutiny reveals that the density of
state show a singularity at $E=\mu -2t$. That is%
\begin{equation}
\rho (E)=\frac{1}{\sqrt{4t^{2}-(\mu -E)^{2}}}.
\end{equation}%
If $\mu =\mu _{c}$, the density of state is singular at $E=0$. Such a
singularity makes a huge number of low-lying excitations contribute to the
entropy, hence suppresses the temperature significantly at the transition
point.

\section{Experimental detection of quantum criticality}

\label{sec:diss}

\begin{figure}[tbp]
\includegraphics[width=8cm]{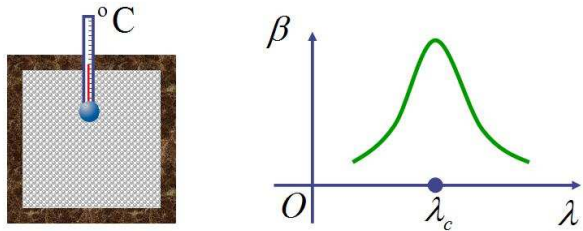}
\caption{(Color online) A sketch of the adiabatic detection of quantum
critical point. Left: a quantum many-body system is adiabatically isolated
from the environment except for a thermometer that can measure the
temperature of the system but is assume to have no obvious interference to
the system. Right: the expected behavior of the adiabatic inverse
temperature around the critical point.}
\label{fig:sketch.eps}
\end{figure}

Armed with the above motivation, we now try to explore the experimental
possibility to detect quantum criticality in quantum many-body systems. A
sketch of our scheme is shown in Fig. \ref{fig:sketch.eps}. We assume a
quantum many-body system described by the Hamiltonian [Eq. (\ref%
{eq:Hamiltonian})] is isolated adiabatically (see the left of Fig. \ref%
{fig:sketch.eps}). A thermometer is used to detect the temperature of the
system. Moreover, we assume that the thermometer is small enough that the
heat transfer between the thermometer and the system can be neglected. Now
an experimentalist first prepare the system in the gapped region and try to
cool the system down to a temperature $\beta _{0}$ that is comparable to the
energy gap $\Delta E(\lambda )$. Secondly, let the system adiabatically
evolve to the another phase. According to the above motivation, the
vanishing gap will make the temperature to reach a minimum at the critical
point, and the adiabatic inverse temperature might be divergent (See the
right of Fig. \ref{fig:sketch.eps} for a sketch).

We believe that to do such an experiment is of great importance both in
physics and information science. Our final conclusion actually is based on
the two theorems emerging from two distinct fields, i.e., quantum
statistical mechanics and quantum information science, respectively. An
experimental observation of a significant enhancement of the adiabatic
inverse temperature around critical points would witness the success of the
combination of the two theorems.

On the other hand, in thermodynamics, for an isolated ideal gas, an
adiabatic cooling occurs when its pressure is decreased adiabatically as it
does work on its surroundings. In our scheme, a low-temperature adiabatic
process in the vicinity of the critical point can also adiabatically cool
the system. Compared with that occurs in the isolated ideal gas, the
adiabatic cooling here is of quantum-like, and not due to doing work on the
surroundings. For instance, for the spin discussed in section \ref%
{sec:warm-up}, the internal energy is%
\begin{equation}
\left\langle E\right\rangle =-h\tanh \left[ \beta \text{sgn}(h)\right]
\end{equation}
where sgn($x$) is the sign function of $x$. Clearly $\left\langle
E\right\rangle$ reaches a maximum at $h=0$ if $\beta $sgn$(h)=$const, so the
internal energy increases as the system tends to the critical point. We
interpret it due to that the ground-state energy increases a lot during such
a process. The surrounding should still do work to the system. The
discrepancy can be solved if we regard the ground-state energy as reference
potential energy. The the internal energy becomes
\begin{equation}
\left\langle E-E_{0}\right\rangle =-h\tanh \left[ \beta \text{sgn}(h)\right]
+h\text{sgn}(h)
\end{equation}
which reaches a minimum at $h=0$. Therefore, if we extract the ground-state
energy from the internal energy, the system does ``do work" to surroundings.
In any case, a potential issue might be that if such a cooling technique is
experimentally useful to reach ultra-low temperatures.

\section{Summary and discussions}

\label{sec:sum}

In summary, we have found that the quantum criticality can be detected by a
low-temperature adiabatic process. The combination of the entropy majorization
theorem in quantum information science and the adiabatic theorem in statistical
physics manifests that the adiabatic inverse temperature might become singular
at the critical point due to the significant change of the low-lying energy
spectra around the critical point. Such a straightforward motivation has been
illustrated by two simple quantum phase transitions occurring in the ground
state of the one-dimensional transverse-field Ising model and spinless fermion
system respectively. Since the temperature is an experimentally measurable
quantity, our work, therefore, provides an operatable scheme to detect quantum
criticality in view of both quantum information science and statistical
physics.

Meanwhile, though we give only two examples to show that the adiabatic
inverse temperature might diverges around the critical point. Entropy
majorization theorem, as discussed in section \ref{sec:major}, gives a clear
picture that the temperature should be significantly suppressed as the
density of state is enhanced in the adiabatic process. Therefore, such a
phenomenon is believed to be universal in all quantum phase transitions.

However, we would like to emphasize here again that starting temperature
should be low enough. For a quantum phase transition from a gapped phase to
another gapped phase, the temperature should be comparably lower than the
gap. While for the quantum phase transition discussed in section \ref%
{sec:fermions}, since the transition is related to the singularity in the
density of the state, the starting temperature in the non-polarized phase
should not be larger enough to excite the state around the singular point.

This work is supported by the Earmarked Grant Research from the Research Grants
Council of HKSAR, China (Project No. CUHK 400807). We thank Dong Yang for
helpful discussions. We are grateful for the hospitality of the Erwin Schr%
\"{o}dinger International Institute for Mathematical Physics at the University
of Vienna, where the work is finalized.

\end{document}